\documentstyle[12pt]{article}
\newlength{\extraspace}
\newlength{\extraspaces}
\begin{document}

\thispagestyle{empty}

\hfill \parbox{3.5cm}{LMU-TPW-99/16 \\ MPI-PhT/99-36 \\ SIT-LP-99/08 \\ (final version)}
\vspace*{1cm}
\begin{center}
{\bf SUPERSYMMETRIC STRUCTURE OF SPACETIME AND MATTRE  \\
--SUPERON-GRAVITON MODEL--} \\[20mm]
{Kazunari SHIMA} \\[2mm]
{\em Sektion Physik der Ludwig-Maximilians-Universit\"at}\\
{\em Theresienstr. 37, D-80333 M\"unchen, Germany}\\[2mm]
{\em Max-Planck-Institut f\"ur Physik}\\
{\em (Werner-Heisenberg-Institut)}\\
{\em F\"ohringer Ring 6, D-80805 M\"unchen, Germany}\\[2mm]
{\em Laboratory of Physics,  Saitama Institute of Technology}\footnote{Permanent address/e-mail: shima@sit.ac.jp}\\
{\em Okabe-machi, Saitama 369-0293, Japan}\\[2mm]
{November  2000}\\[15mm]


\begin{abstract}
A unified description of spacetime and matter is proposed by using 
a single irreducible representation of SO(10) Super-Poincar\'e algebra(SO(10)SPA). 
All (observed) elementary particles except the graviton are the (massless) eigenstates of SO10)SPA 
composed of fundamental Nambu-Goldstone fermions with spin 1/2, ${\em superons}$ associating with 
the spontaneous breakdown of the supertranslation of the spacetime.
The systematic investigations of the standard model(SM) and SU(5)GUT by using superon diagrams 
may reveal the generation structure, the stability of the proton, $K^{0}- \overline K^{0}$, 
$B^{0}- \overline B^{0}$ and $D^{0}- \overline D^{0}$  mixings, CP-violation, the atmospheric 
and solar neutrino deficits and  the absence of the electroweak lepton-flavor-mixing. 
The fundamental action of the  superon-graviton model(SGM) for spacetime and matter is proposed, 
which is invariant under a new supersymmetry.   \\

\end{abstract}
\end{center}

\newpage

The supersymmetry (SUSY)[1] gives a natural framework to unify spacetime and matter and 
is  expected to give the solutions to  many unsolved  fundamental problems 
in the SM and GUTs, for example,
the origin of the generation structure
of quarks and leptons, the absence of the electroweak mixings  among
the lepton generations, the stability of the proton and the accomodation of the
gravitational interaction, ...etc.
Unfortunately as shown by Gell-Mann[2] by the group theoretical arguments, SO(8) maximally extended supergravity
theory (SUGRA) is too small to accommodate all observed particles as the elementary fields.
It is well known within the S-matrix arguments of the local gauge field theory 
that SO(N) SUGRA with ${N>8}$ does not exist 
due to so called the no-go theorem[3] for the high-spin($>2$) ${\em massless}$ ${\em elementary}$ fields.
However, we think that from the viewpoint of simplicity and beauty of nature
and also from the viewpoint of a probable mean to recognize the Planck scale
physics in the case of spacetime having a certain boundary(i.e. a boundary condition),
it is interesting to attempt the accomodation of all observed  particles in a single
irreducible representation of a certain group(algebra).
Also the no-go theorem does not exclude the possibility that the fundamental action, if it exists,
posesses the high-spin degrees of freedom not as the elementary fields but as some ${\em composite}$
eigenstates of a certain symmetry (algebra) of the fundamental action. 
In this letter we would like to pursue the possibility of this scenario.
Furthermore despite the advocated success of the superstring theory, I think that the physics
at the Planck scale is still an  unknown exciting problem to be challenged and allows various attempts.       \\
In Ref.[4], by the group theoretical arguments we have shown that among all single irreducible representations of 
all SO(N) extended super-Poincar\'e(SP) symmetries, the massless irreducible
representations of SO(10) SP algebra(SPA) is the only one that accomodates minimally all observed particles 
including the graviton. However the fundamental theory was left unknown.                                            \\
In Ref.[5], we have shown that  SO(10) SPA  for the massless
irreducible representation indicates by itself a certain compositeness of all
elementary particles except the graviton. We have identified the fundamental
constituents with Nambu-Goldstone(N-G) fermions  ${\em superons}$ corresponding to the
spontaneous breakdown of the supertranslation of the spacetime and proposed a fundamental
theory ${\em superon}$-${\em graviton}$ ${\em model}$(SGM) by extending Volkov-Akulov
nonlinear SUSY action to the curved spacetime.                                           \par
 In this article, with a brief summary of ref.[4][5] for the self-contained arguments
we study the fundamental action of SGM further and show that it is invariant under a new supersymmetry, 
which favours SGM scenario. 
Furthermore we will see that SGM does  not suffer from the no-go theorem for the massless high-spin 
${\em elementary}$  field.
Because the high-spin($>2$) massless degrees of freedom of the fundamental action are accomodated 
not as the dangerous massless elementary fields but as the (massless)  eigenstates composed of N-G fermions.    \\
In ref [4][5] by noting that 10 generators $Q^{N}(N=1,2,..,10)$ of SO(10) SPA are the
fundamental represemtations of SO(10) internal symmetry  and
$SO(10) \supset SU(5) \supset SU(3) \times SU(2) \times U(1)$
we have decomposed 10 generators $Q^{N}$ of SO(10) SPA as follows with
respect to $SO(10) \supset SU(5) \supset SU(3) \times SU(2) \times U(1)$
\begin{eqnarray}\underline{10} & = & \underline 5
+ \underline 5^{*}  \nonumber \\
& = & \{(\underline 3, \underline 1;-{1 \over 3},-{1 \over 3},-{1 \over 3}) +
 (\underline 1, \underline 2;1, 0)\} +
 \{(\underline 3^{*}, \underline 1;{1 \over 3},{1 \over 3},{1 \over 3}) +
 (\underline 1, \underline 2^{*};-1,0)\},
\end{eqnarray}
where we have specified (${\underline{SU(3)},\underline{SU(2)}}$; electric charges ).
To obtain a smaller single irreducible representation we have studied the massless
representation.
For massless  case the little algebra for the
supercharges in  the light-cone frame $P_{\mu}=\epsilon(1,0,0,1)$  becomes
after a suitable rescaling
\begin{equation}
\{ Q_{\alpha}^{M}, Q_{\beta}^{N} \}
=\{ \bar{Q}_{\dot\alpha}^{M}, \bar{Q}_{\dot\beta}^{N} \}=0, \quad
\{Q_{\alpha}^{M},\bar{Q}_{\dot\beta}^{N}\}
={\delta}_{{\alpha}1}{\delta}_{{\dot\beta}{\dot1}}{\delta}^{MN},
\end{equation}
where $\alpha,\beta=1,2$ and $M,N=1,2,...5$.
Note that the spinor charges
$Q_{1}^{M}$, $\bar{Q}_{\dot 1}^{M}$
satisfy the algebra of annihilation and creation operators of the massless fermions respectively
and can be used to construct a finite dimensional supersymmetric Fock space
with positive metric.
We identify the graviton with the Clifford vacuum $\mid\Omega(\psi)\rangle$
(SO(10) singlet but not necessarily the lowest energy state) satisfying
$Q_{\alpha}^{M} \mid\Omega(\psi)\rangle=0$, which generates automatically the
adjoint representation of SO(10) at helicity ${\pm1}$ state. This identification is physically natural,
for only the graviton does not distinguish between  bosons and  fermions.
By performing the ordinary procedures we obtain  $2\cdot2^{10}$ dimensional irreducible representation of
the little algebra (2) of SO(10) SPA as follows[4]:     \\
$\Bigl[\underline{1}(+2), \underline{10}(+{3 \over 2}),
\underline{45}(+1), \underline{120}(+{1 \over 2}),
\underline{210}(0),
\underline{252}(-{1 \over 2}),
\underline{210}(-1),
\underline{120}(-{3 \over 2}),
\underline{45}(-2), \\
\underline{10}(-{5 \over 2}), \underline{1}(-3)\Bigr]
+ \Bigl[ \mbox{CPT-conjugate} \Bigr]$,                          \\
where $\underline{d}(\psi)$  represents
SO(10) dimension $\underline{d}$ and the helicity $\psi$.                      \\
By noting that the helicities of these  states are
automatically determined by SO(10) SPA in the light-cone and that
$Q_{1}^{M}$ and  $\bar{Q}_{\dot{1}}^{M}$
satisfy the algebra of the annihilation and the creation operators for the
massless spin ${1 \over 2}$ particle, we speculate boldly that
these massless states spanned upon the  Clifford vacuum  $\mid\Omega(\pm2)\rangle$
are the $\it{relativistic(gravitational)}$ massless  eigenstates composed of the
fundamental massless object $Q^{N}$, $\it{superon}$ with spin ${1 \over 2}$.
Therefore we regard (1) as $\it{a}$ ${superon}$-${quintet}$
and $\it{an}$ ${antisuperon}$-${quintet}$.
These ${\em massless}$ states may not be  necessarily the bound nor resonance states
but the eigenstates of spacetime and matter with SO(10) SP symmetric structure,
because they are as shown later the composites of ${\em massless}$ N-G fermion superons and
correspond merely to all possible nontrivial combinations of the multiplications of
the spinor charges(i.e. generators) of SO(10) SP symmetry.
The unfamiliar identification of the generators of SO(10) SP algebra with the
fundamental objects is discussed later.
Now we envisage the Planck scale physics as follows:        \\
At(above) the Planck energy scale spacetime and  matter have the
structure described by SO(10) SPA, where the gravity dominates and creates
the superon-quintet and the antisuperon-quintet pair( not a single
particle-antiparticle pair ) from the vacuum in such a way as superon-composite
massless states, i.e. all possible nontrivial combinations
of the superons, span the massless irreducible representations
of SO(10) SPA, i.e. the eigenstates of spacetime and matter.                  \par
From the viewpoints of the superon-quintet model(SQM) for matter
we can study more concretely the physical meaning of the results obtained in Ref.[4][5]. 
Hereafter we adopt the following symbols for superons:
For the superon-quintet 
$\underline 5$ = $\Bigr[ $(\underline 3, \underline 1;-${1 \over 3}$,-${1 \over 3}$,
-${1 \over 3}$)$,$(\underline 1, \underline 2;1, 0)$ \Bigl]$[5], we use
\begin{equation}
\Bigr[ Q_{a}(a=1,2,3) ,Q_{m}(m=4,5)\Bigl]
\end{equation}
\noindent and for the antisuperon-quintet 
$\underline 5^{*}$ = $\Bigr[ (\underline 3^{*}, \underline 1;+{1 \over 3},+{1 \over 3},
+{1 \over 3})$,$(\underline 1, \underline 2^{*};-1, 0) \Bigl]$[5], we use
\begin{equation}
\Bigr[Q_{a}^{*}(a=1,2,3),Q_{m}^{*}(m=4,5)\Bigl],
\end{equation}
where ${a=1,2,3}$ and ${m=4,5}$ represent the color and electroweak components of
superons respectively.
Interestingly our model needs only the superon-quintet who have 
the same quantum numbers  as the fundamental matter multiplet ${\underline 5}$ of SU(5) GUT[6]: 
For the color component superons
\begin{equation}
\Bigr[Q_{a},{\underline 3}=\{a\}, \underline 1 , 0, -{2 \over 3}, 0 \Bigl]
\end{equation}
and for the electroweak component superons
\begin{equation}
\Bigr[Q_{m},{\underline 1}, {\underline 2}=\{m\}, {{(-)^{m}}{1 \over 2}},0,-1 \Bigl],
\end{equation}
where we have shown $\Bigr[Q_{N}, {\underline {SU(3)}}_{color},{\underline {SU(2)}}_{weak}, I_{z},
B(baryon\ number),                 \\
L(lepton\ number)\Bigl]$. Note that the Gell-Mann--Nishijima relation is satisfied by the quantum numbers
of the superon.     \\
\begin{equation}
Q_{e}=I_{z} + {1 \over 2}(B-L).
\end{equation}
Accordingly all the the states  are specified explicitly with respect to
( SU(3), SU(2); electric charges ).
Suppose that through  the symmetry breaking
[SO(10) SPA upon the Clifford vacuum]
$\longrightarrow$ [ \ $SU(3) \times SU(2) \times U(1)$ ]
$\longrightarrow$ [ \ $SU(3) \times U(1)$ ] which is
discussed later, many of the lower helicity
$( \pm{3 \over 2}, \pm1, \pm{1 \over 2}, 0 )$ states
of SO(10) SPA are converted to the longitudinal components of the higher spin
massive states of PA in  $SU(3) \times SU(2) \times U(1)$ invariant way
and others remain massless.
That is, all unnecessary (for SM) higher helicity states become massive by eating
the lower helicity states in $SU(3) \times SU(2) \times U(1)$ invariant way.
We have carried out the recombinations of the states (corresponding to the superHiggs mechanism
and/or to the diagonlizations of the mass terms of the high-spin fields)
among $2 \cdot 2^{10}$ helicity states and
found surprisingly all massless states necessary  and sufficient for
the SM with three generations of quarks and leotons appear in the surviving
massless states (therefore, no sterile neutrinos). At least one new spin ${3 \over 2}$ lepton-type doublet
$( \nu_{\Gamma}, \Gamma^{-} )$ with the mass of the electroweak scale is
predicted[4]. \\
As for the assignments of observed particles, we take for simplicity   
the following left-right symmetric assignment for quarks and  leptons
by using the conjugate representations  naively , i.e.
$( \nu_{l}, \it{l}^{-} )_{R}= (\bar\nu_{l}, \it{l}^{+} )_{L}$, etc[5].
Furthermore as for the generation assignments we assume simply that the states
with more (color-) superons turn to acquiring larger masses in the low energy and no mixings among genarations.
The surviving massless states identified with  SM(GUT) are  as follows.                             \\
For three generations of leptons
[$({\nu}_{e}, e)$,  $({\nu}_{\mu}, \mu)$,  $({\nu}_{\tau}, \tau)$], we take
\begin{equation}
\Bigr[(Q_{m}{\varepsilon}_{ln}Q_{l}^{*}Q_{n}^{*}),
(Q_{m}{\varepsilon}_{ln}Q_{l}^{*}Q_{n}^{*}Q_{a}Q_{a}^{*}),
(Q_{a}Q_{a}^{*}Q_{b}Q_{b}^{*}Q_{m}^{*})\Bigl]
\end{equation}
\noindent and the conjugate states respectively.
\\
For three generations of quarks [$( u, d )$, $( c, s )$, $( t, b )$],
we have ${\em uniquely}$
\begin{equation}
\Bigr[({\varepsilon}_{abc}Q_{b}^{*}Q_{c}^{*}Q_{m}^{*}),
({\varepsilon}_{abc}Q_{b}^{*}Q_{c}^{*}Q_{l}
{\varepsilon}_{mn}Q_{m}^{*}Q_{n}^{*}),
({\varepsilon}_{abc}Q_{a}^{*}Q_{b}^{*}Q_{c}^{*}Q_{d}Q_{m}^{*})\Bigl]
\end{equation}
\noindent and the conjugate states respectively. \\
For $SU(2) \times U(1)$ gauge bosons [ ${{W}^{+},\ Z,\ \gamma,\ {W}^{-}}$], we have from
${{\underline 2} \times {\underline 2^{*}}}$  \\
${[Q_{4}Q_{5}^{*}, {1 \over \sqrt{2}}( Q_{4}Q_{4}^{*} \pm Q_{5}Q_{5}^{*}),
Q_{5}Q_{4}^{*}]}$. \\
For $SU(3)$ color-octet gluons [${G^{a}(a=1,2,..,8)}$], we have \\
${[Q_{1}Q_{3}^{*},Q_{2}Q_{3}^{*},-Q_{1}Q_{2}^{*},
{1 \over \sqrt{2}}(Q_{1}Q_{1}^{*}-Q_{2}Q_{2}^{*}), Q_{2}Q_{1}^{*},
{1 \over \sqrt{6}}(2Q_{3}Q_{3}^{*}-Q_{2}Q_{2}^{*}-Q_{1}Q_{1}^{*})}$,  \\
${-Q_{3}Q_{2}^{*},Q_{3}Q_{1}^{*}]}$.  \\
For [$SU(2)$ Higgs Boson], we have ${[{\varepsilon}_{abc}Q_{a}Q_{b}Q_{c}Q_{m}]}$
and the conjugate state.  \\
For [$( X,Y )$] leptoquark bosons in GUTs,
we have $[Q_{a}^{*}Q_{m}]$  and  the conjugate state.   \\
For a color- and SU(2)-singlet neutral gauge boson from
${{\underline 3} \times {\underline 3^{*}}}$ (which we call simply S  boson
to represent the singlet) we have ${Q_{a}Q_{a}^{*}}$. \\
The specification of $(X,Y)$ gauge boson is important for the proton decay in SU(5) GUT.
The specification of  S boson is interesting as an additional U(1) expected
from the branching rule $SO(10) \supset SU(5) \times U(1) \supset SM \times U(1)$ for SUSY SM.
As shown later S boson plays crucial roles in the process concerning the third
generation of quarks and leptons.
For the vector gauge bosons we have considered only two-superons states
$\underline{45}$ of the adjoint representation of SO(10) SPA.         \\   
Now in order to see the physical implications of superon-quintet model(SQM) for matter
we try to interpret the Feynman diagrams of SM and GUTs in terms of the Feynman
diagrams of SQM.
The superon line Feynman diagrm is obtained by replacing the
single line in the Feynman diagram of the unified gauge models by the corresponding multiple superon lines.
To translate the vertex of the Feynman diagram of
the unified gauge models into that of SQM, we assume that
{\em the pair creation and the pair annihilation of the superon-quintet and the antisuperon-quintet
within a single state for a quark, a lepton and a (gauge) boson (,i.e. within a single SO(10) SPA state)
are rigorously forbidden.}                    \\
This rule( analogous to OZI-rule of the quark model )  seems natural
because each state is an irreducible representation
of SO(10) SPA and is prohibitted from the decay without any remnants,
i.e.  without the interaction between the superons contained
in the different states. As discussed later, this my be  equivalent to the absence of the excited states
of quarks, leptons, gauge bosons, .. etc.
Here we just mention that all the states necessary for  SM and GUTs
with three genarations of quarks and leptons appear up to  the five-superons states
i.e. one half of the whole eigenstates of SO(10) SPA. 
This may suggest a certain unknown new mechanism which produces a large mass splitting between these states 
and the others.                                                    \par
Now the translation is unique and straightforward.                 
We see that at the Yukawa coupling vertex of SQM the observed quark
(the observed lepton) interacting  with the Higgs boson couples to a new quark(a new lepton)
which is exotic with respect to SU(2) and/or spin.
Then the Yukawa coupling of SM(GUTs) appears effectively only in the higher orders of
the Yukawa couplings of SQM, which gives the Yukawa coupling
of SM(GUTs) a small factor of the order of the inverse of the large masses
of the exotic quark and lepton.
This mechanism  may give a clue to understand the structure of
the CKM mixing matrix for the quark sector but
may be dangerous so far for the lepton sector because of the disastrous electroweak 
lepton-flavor violation by the lepton mixing.
However we find that at every vertex of the gauge coupling there is a stringent
selection rule for generations which is characteristic only to SQM,
for each generation is identical only with respect to
${SU(3)\times SU(2) \times U(1)}$ quantum numbers but has another superon content
corresponding to the flavor quantum number.
This selection rule is the matching of the superons, i.e. the superon number
conservation, at the (gauge) coupling vertex. For the quark sector, surprisingly,
the selection rule respects the CKM mixing of the Yukawa coupling sector and maintains
the successful electroweak gauge current structure of SM (except the third generations).
While for the lepton sector, remarkably the selection rule forbids 
the lepton-flavor-changing electroweak currents between lepton generations
at the tree level and realizes the success of SM. That is, the  absence of
the electroweak lepton-flavor-mixing at the tree level is derived by the superon pictures. \\
As a few examples of the gauge  interactions and the selection rule at the
gauge coupling vertex we demonstrate  the following typical processes, i.e.
\noindent $(i)$ ${\beta}$ decay[5]:
$n \longrightarrow p + e^{-} + \bar{\nu_{e}}$,\quad
$(ii)$ ${\pi}^{0} \longrightarrow 2{\gamma}$[5],
$(iii)$ the proton decay of GUT[5]: $p \longrightarrow e^{+} + {\pi}^{0}$,
$(iv)$ the poroton decay of SUSY GUT: ${p \longrightarrow  K^{+} +  \bar{\nu}}$,
$(v)$  a flavor changing neutral current process(FCNC)[5]:
$K^{+} \longrightarrow \pi^{+} + \nu_{e} + \bar\nu_{e}$
and
$(vi)$ an advocated typical process of the new physics beyond the SM[5]:
$\mu \longrightarrow e + \gamma$. \\
\noindent
For the processes $(i)$ and $(ii)$ we can draw the corresponding
tree-like superon line diagrams easily, which is tree-like. 
For the process $(iii)$ we examine the Feynman diagrams for the proton decay
of GUTs and find that the corresponding superon line diagrams
do not exist due to the selection rule, i.e. the mismatch of the superons
contained in the quarks($u$ and $d$)  and the gauge bosons($X$ and $Y$)
at the gauge coupling vertices.
This means that irrespective of the massses of the gauge bosons the proton is stable
at the tree level against $p \longrightarrow e^{+} + {\pi}^{0}$.
$(iv)$ We just mention that the proton decay ${p \longrightarrow  K^{+} +  \bar{\nu}}$,
which is the dominant decay mode of SUSY GUT represented by the box-type (gaugino-Higgsino exchange)
higher order (dimension 5) Feynmann diagram, is forbidden similarly by the selection rule
(not by R-parity).
For FCNC process $(v)$ the penguin-type and the box-type superon line
diagrams are to be studied corresponding to the penguin- and box-Feynmann
diagrams for
$K^{+} \longrightarrow \pi^{+} + \nu_{e} + \bar\nu_{e}$ of GUTs.
Remarkablely the superon line diagrams which have only the $u$ and  $c$ quarks for the
internal quark line exist due to the selection rule and  GIM mechanism is reproduced.
This is the indication of the strong suppression of the FCNC process,
$K^{+} \longrightarrow \pi^{+} + \nu_{e} + \bar\nu_{e}$. This simple mechanism
may hold in general for FCNC processes.
For the process $(vi)$ the corresponding tree-like superon line diagram
does not exist due to the selection rule at the gauge coupling vertex,
i.e.  $\mu \longrightarrow e + \gamma$ decay mode is
absent at the tree-level in the superon (composite) model.  The process
$\tau \longrightarrow e(\mu) + \gamma$ is suppressed similarly. \\
\noindent As for the CP-violation the mixing
$K^{0}$-$\overline{K^{0}}$ is natural in SQM,
for remarkably the superon contents of $K^{0}$ and $\overline{K^{0}}$ are the same
but in the different combinations distinguished only by the interactions, i.e.
$K^{0}$ = $d \overline{s}$ =$({\varepsilon}_{abc}Q_{b}^{*}Q_{c}^{*}Q_{4}^{*})
({\varepsilon}_{ade}Q_{d}Q_{e}Q_{5}^{*}Q_{4}Q_{5})$ and
$\overline{K^{0}}$= $\overline{d}s$ = $({\varepsilon}_{abc}Q_{b}Q_{c}Q_{4})
({\varepsilon}_{ade}Q_{d}^{*}Q_{e}^{*}Q_{5}Q_{4}^{*}Q_{5}^{*})$.
GIM mechanism works for the superon picture of $K^{0}$-$\overline{K^{0}}$
mixing box diagram of SM, but remarkably $t$ quark (the third generation) decouples
due to the selection rule at the gauge coupling vertices.
However in SQM there is another higher order box doiagram contributing to
$K^{0}$-$\overline{K^{0}}$ mixing amplitude, where  $t$ quark  and S gauge boson
emitted by the transition ${( u , d ) \leftrightarrow ( t , b )}$
play crucial roles besides W boson.
The relative phase of these two amplitudes may be an  origin of CP-violation
in the neutral K-meson decay. Interestingly, the third generation of quarks is needed
for CP-violation in this different context.
The mixings $B^{0}$-$\overline{B^{0}}$ and  $D^{0}$-$\overline{D^{0}}$  are natural
in the same sense, for example,
 $B^{0}$ = $d \overline{b}$ =$({\varepsilon}_{abc}Q_{b}^{*}Q_{c}^{*}Q_{4}^{*})
({\varepsilon}_{def}Q_{d}Q_{e}Q_{f}Q_{a}^{*}Q_{4})$ and
$\overline{B^{0}}$= $\overline{d}b$ = $({\varepsilon}_{abc}Q_{b}Q_{c}Q_{4})
({\varepsilon}_{def}Q_{d}^{*}Q_{e}^{*}Q_{f}^{*}Q_{a}Q_{4}^{*})$.
But the preliminary analyses suggest the similar new mechanisms
for mixing and CP violation characteristic of the SQM.
As for the charmless nonleotonic B decay[7]  in SQM
the transition ${( t , b ) \leftrightarrow ( c , s )}$
occurrs not at the tree level of the weak charged current but at the higher orders of
the gauge couplings due to the selection rule for the quark sector,
where  the transition ${( t , b ) \leftrightarrow ( c , s )}$  is achieved by
the emissions of S boson and W boson  and may be an origin of the excess of
the charmless(or the suppression of the charm mode) nonleotonic B decay[7].     \\
Here we should just mention an alternative  assignment of the lepton sector. 
SQM posesses originally one more (forth) electroweak lepton-doublet state composed of three superons 
$(Q_{a}Q_{a}^{*}Q_{m}^{*})$. This state is absorbed by spin 3/2  and absent in the above assignment (8), 
which  gives a systematic(universal) description of the recently observed mixings(oscillations) of the lepton sector 
${\nu_{e} \longleftrightarrow \nu_{\mu} \longleftrightarrow \nu_{\tau}}$ 
by the emission of a scalar particle $(Q_{a}Q_{a}^{*}Q_{m}Q_{m}^{*})$, which is beyond the SM. 
If the second generation of (8) is replaced by $(Q_{a}Q_{a}^{*}Q_{m}^{*})$, the transition
${\em only}$ ${{\nu}_{\mu} \leftrightarrow {\nu}_{\tau}}$ (i.e. the transiton
between the second and the third genertion) is induced by the S gauge boson 
at the tree level due to the selection rule, which may explain simply and naturally  the
${{\nu}_{\mu}}$ deficit problem of the atomospheric neutrino[8] besides the neutrino oscillations.    \\
Next we just mention the excited states of quarks, leptons and  gauge bosons.
As stated before  these particles corresponding to the (massless) eigenstates of SO(10) SP symmetry
do not have the (low energy) excited states in SQM,
because each particle is a single (massless) eigenstates of SO(10) SP symmetry
composed of superons and transits to ${\it another}$ ${\it eigenstate}$ through the
interaction, i.e. through the absorption or the emission of superons
(,i.e. eigenstates).
Does this explain the absence of the low energy excited states of the observed quarks, leptons and gauge bosons,
even if they are composites?                                                    \par
Finally we consider the fundamental theory of SGM
for supersymmetric spacetime and matter.
In carrying through the canonical quantization of the elementary
(N-G) spinor field $\psi(x)$ of Volkov-Akulov model[9] of
the nonlinear SUSY(NL SUSY), we have shown that the supercharges $Q$ given
by the supercurrents
\begin{equation}
J^{\mu}(x)={1 \over i}\sigma^{\mu}\psi(x)
-\kappa \{ \mbox{the higher order terms of $\kappa$, $\psi(x)$ and }\  \partial\psi(x) \}
\end{equation}
obtained by the ordinary Noether procedures can satisfy the super-Poincar\'e algebra
at the cnonically quantized level[10].
(10) means the field-current identity between the elementary N-G spinor field
$\psi(x)$ and the supercurrent, which justifies our bold assumption
that the generator(supercharge) $Q^{N}$ (N=1,2,..10) of SO(10) SPA in the light-cone frame represents
the fundamental massless particle, $superon$ $with$ $spin$ ${1 \over 2}$.
That is, supersymmetry indicates the existence of the supercharged superons.
Therefore we speculate that the fundamental theory of SQM for matter is SO(10) NLSUSY
and that the fundamental theory of SGM for spacetime and matter
at(above) the Planck scale is SO(10) NL SUSY in the
curved spacetime(corresponding  to the Clifford vacuum $\mid\Omega(\pm 2)\rangle$).
We can regard that all the massless
(helicity)-states of SO(10) SPA including the observed
quarks, leptons and gauge bosons except the graviton are the relativistic
${\it gravity}$-${\it induced}$ composite massless eigenstates composed of
massless N-G fermion $superons$ originating from the spontaneous breakdown of
the supertranslation.                                      \\
SGM may indicate that the old dream of constructing the relativistic composite (quark) model of
matter may be realized as eigenstates of SO(10) SPA at the superon level.
It is interesting that the field theory of the composite model of quarks and leptons
based upon NL SUSY was  already challenged long time ago[11].                       \par

Now we consider further a fundamental theory of SGM and reinvestigate the action given in Ref.[5]. 
We extend the arguments of Volkov-Akulov[9] to the curved spacetime, 
where NL SUSY SL(2C) degrees of freedom (i.e. the coset space coordinates representing N-G fermions) 
in addition to  Lorentz SO(3,1) coordinates are embeded at every curved spacetime point with GL(4R) invariance. 
As discussed later  we can define a new tetrad and a new metric tensor in the abovementioned curved spacetime  and 
obtain the following Lagrangian as the fudamental theory of SGM for  spacetime and matter.

\begin{equation}
L=-{c^{3} \over 16{\pi}G}\vert w \vert(\Omega + \Lambda ),
\end{equation}
\begin{equation}
\vert w \vert=det{w_{a}}^{\mu}=det({e_{a}}^{\mu}+\kappa {t_{a}}^{\mu}),  \quad
{t_{a}}^{\mu}={1 \over 2i}\sum_{j=1}^{10}(\bar{\psi}^{j}\gamma_{a}
\partial^{\mu}{\psi}^{j}
- \partial^{\mu}{\bar{\psi}^{j}}\gamma_{a}{\psi}^{j}),
\end{equation} 
where $\kappa$ is a four dimensional fundamental volume of superspace,  
${e_{a}}^{\mu}(x)$ is the vierbein of Einstein general relativity theory(EGRT) and 
$\Lambda$ is a probable cosmological constant. 
$\Omega$ is a new scalar curvature analogous to the Ricci scalar curvature $R$ of EGRT. The explicit 
expression of $\Omega$ is obtained  by just replacing ${e_{a}}^{\mu}(x)$  by ${w_{a}}^{\mu}(x)$ in Ricci scalar $R$. 
Therefore the lowest order of $\kappa$, i.e. the superonless vacuum, of the action (11) gives the Einstein-Hilbert 
action of general relativity. 
The action  (11) is invariant at least under  GL(4R), local Lorentz, global SO(10) and the following new supersymmetry
\begin{equation}
\delta \psi^{i}(x) = \zeta^{i} + i \kappa (\bar{\zeta}^{j}{\gamma}^{\rho}\psi^{j}(x)) \partial_{\rho}\psi^{i}(x),
\end{equation} 
\begin{equation}
\delta {e^{a}}_{\mu}(x) = i \kappa (\bar{\zeta}^{j}{\gamma}^{\rho}\psi^{j}(x))\partial_{[\rho} {e^{a}}_{\mu]}(x),
\end{equation} 
where $\zeta^{i}, (i=1,..10)$ is a constant spinor and  $\partial_{[\rho} {e^{a}}_{\mu]}(x) = 
\partial_{\rho}{e^{a}}_{\mu}-\partial_{\mu}{e^{a}}_{\rho}$. 

These results can be understood intuitively by observing that 
${w_{a}}^{\mu}(x) ={e_{a}}^{\mu}(x)+\kappa {t_{a}}^{\mu}(x)$ in (12) defined by 
$\omega_{a}={w_{a}}^{\mu}dx_{\mu}$, where $\omega_{a}$ is the NL SUSY invariant differential forms of 
Volkov-Akulov[9], and  $s^{\mu \nu}(x) \equiv {w_{a}}^{\mu}(x) w^{{a}{\nu}}(x)$ are formally  
a new vierbein and a new metric tensor in the abovementioned curved spacetime. 
In fact, it is not difficult to show the similarity of ${w_{a}}^{\mu}(x)$ and 
$s^{\mu \nu}(x)$ to ${e_{a}}^{\mu}(x)$ and $g^{\mu\nu}(x)$, i.e., 
${w_{a}}^{\mu}(x)$ and $s^{\mu \nu}(x)$ are invertible, 
${w_{a}}^{\mu} w_{{b}{\mu}} = \eta_{ab}$,  $s_{\mu \nu}{w_{a}}^{\mu} {w_{b}}^{\mu}= \eta_{ab}$, ..etc. 
and the following transformations of ${w_{a}}^{\mu}(x)$ and $s_{\mu\nu}(x)$ under (13) and  (14) 
\begin{equation}
\delta_{\zeta} {w^{a}}_{\mu} = \xi^{\nu} \partial_{\nu}{w^{a}}_{\mu} + \partial_{\mu} \xi^{\nu} {w^{a}}_{\nu}, 
\end{equation} 
\begin{equation}
\delta_{\zeta} s_{\mu\nu} = \xi^{\kappa} \partial_{\kappa}s_{\mu\nu} +  
\partial_{\mu} \xi^{\kappa} s_{\kappa\nu} 
+ \partial_{\nu} \xi^{\kappa} s_{\mu\kappa}, 
\end{equation} 
where  $\xi^{\rho}=i \kappa (\bar{\zeta}^{j}{\gamma}^{\rho}\psi^{j}(x))$.
Therefore the similar arguments to EGRT in Riemann space can be carried out straightforwadly 
by using  $s^{\mu \nu}(x)$ (or ${w_{a}}^{\mu}(x)$) in stead of  $g_{\mu\nu}(x)$ (or ${e_{a}}^{\mu}(x)$), 
which leads to (11) manifestly invariant at least under the above mentioned symmetries. 
The commutators of two new supersymmetry transformations  on $\psi(x)$ and  ${e_{a}}^{\mu}(x)$ 
are 
\begin{equation}
[\delta_{\zeta_{1}}, \delta_{\zeta_{2}}] \psi^{i} = 
\{2i \kappa ({\bar{\zeta}^{j}}_{2}{\gamma}^{\mu}{\bar{\zeta}^{j}}_{1}) 
- \xi_{1}^{\rho} \xi_{2}^{\sigma} {e_{a}}^{\mu} (\partial_{[\rho} {e^{a}}_{\sigma]})\} \partial_{\mu} \psi^{i}, 
\end{equation} 
\begin{equation}
[\delta_{\zeta_{1}}, \delta_{\zeta_{2}}] {e^{a}}_{\mu} = 
\{2i \kappa ({\bar{\zeta^{j}}}_{2}{\gamma}^{\rho}{\bar{\zeta}^{j}}_{1}) 
 -  \xi_{1}^{\sigma} \xi_{2}^{\lambda} {e_{c}}^{\rho} (\partial_{[\sigma} {e^{c}}_{\lambda]})\} 
\partial_{[\rho} {e^{a}}_{\mu]}
 -  \partial_{\mu}(  \xi_{1}^{\rho} \xi_{2}^{\sigma} \partial_{[\rho} {e^{a}}_{\sigma]}), 
\end{equation} 
which form a closed algebra[12].                                                  \par
As for the abovementioned (spontaneous) symmetry breaking it is a challenge to
study the structure of the true vacuum of (11). The order of the mass scale of
spontaneous symmetry breaking is given by $\kappa$ and  $\Lambda$.
We should convert the action (11) into the equivalent linear
broken SUSY SO(10) Lagrangian to see clearly the (low energy) mass spectrum of
the particles spanned upon the true vacuum.
The low-energy structure of the linearized broken SUSY Lagrangian should
involve GUTs, at least the SM with three generations.
For carrying through such complicated scenario it is encouraging to note
that the linearlization of such a nonlinear
fermionic system was already carried out explicitly[13][14].
They investigated in detail the conversions between
N=1 NL SUSY(Volkov-Akulov) model and
the equivalent linear (broken) N=1 SUSY Lagrangian.
The extension of the generic and the systematic arguments based on the superspace[14]
may be useful for the linerization  of the N=10 superon model.
We expect that by taking ${non}$-${perturbatively}$ the true vaccum of (11)
the conversion of the NL SUSY action of superons (11) into the linearlized SUSY
action  with the various  spinor fields is achieved, where  the symmetry is
broken spontaneously at the tree level and the bosonic and the fermionic 
high-spin massless composite states appear as ${\it massive}$ high-spin  fields
(which decouple in the low energy provided they are heavy). This may be the only way
to circumvent the no-go theorem in the low energy effective theory.
(The massive high spin fields have no difficulties  so far.)
For carrying out the conversions it is essential to find the (supersymmetric)
constraints ${\it a}$ ${\it la}$ Ro$\breve c$ek[14] which express the fields of
the multiplet of the linear SUSY in terms of N-G fermions.
In our case the symmetry breaking
${[SO(10)NLSUSY] \longrightarrow [SU(m) \times SU(n) \times ...] \supseteq [SM]}$
which makes high-spin fields massive gives the clues for constructing the constraints.
As a possible mechanism of the dynamical mass generation(i.e. symmetry breaking) among
massless states, the condensation of the tensor states in the adjoint representations
$\underline{45}$ composed of eight superons  should be noticed[15].
Because by applying the arguments of strong gravity
to these massless tensor states we can expect that SL(2C) tensor fields
condense and induce the topological(classical) spontaneous symmetry breaking
via the Higgs potential analogue gauge invariant ${\it self}$-interactions[16].
(The gravity is the gauge field of GL(4R) and does not breake spontaneously.)
Also the dynamical mass generation via so called the quantum completion may be relevant.
It is very interesting if we can regard
the yet hypothetical superon-graviton model(SGM) may be for the unified gauge
models(SM and GUTs) what the BCS(electron-phonon) theory is for
the Landau-Ginzburg theory of the superconductivity.
Especially the nonisotropic spontaneous symmetry breaking  of
the superconductivity(,e.g., BCS with s- and d-state electron pair for gappless high 
$T_{c}$, heavy electron, ..etc) and the superfluidity may be suggestive for SGM.     \\
Furthermore considering the fact that the pure Yang-Mills theory allows high spin composites, 
we can expect that SGM (11) regarded formally as the pure GL(4R) theory of 
$s^{\mu \nu}(x)$(${w_{a}}^{\mu}(x)$), 
though noncompact and indefinite, allows similar high spin composites. 
The proof of the gravitaional superon composite states 
is very interesting problem.( Note that to my knowledge hadrons are not yet realized field theoretically 
as the relativistic bound states of quarks by the gluons.)                   \\
Apart from the linearlization of SGM, it is interesting to fit ${\em all}$
the decay data of all observed elementary particles
in terms of the quark model analogue[17]
SO(10) superon current algebra, which may describe the nonlinear superon dynamics
at the short distance of the spacetime and  may give a qualitative test of SQM[4].
Also it is worth studying other ($R \not= L^{*}$) SQM than ${R = L^{*}}$ symmetric
SQM((8) and (9)) for quarks and leptons.                                         \\
The cosmological implications of SGM (11) is also worth studying.
It may give an explanation of the ${\it birth}$ of the universe in terms of
the notion of the spontaneous breakdown of the ultimate SO(10) SP symmetry and its spontaneous breakdown
at(above) the Planck scale, where the SUSY is realized nonlinearly.
SGM (11) describes a pre-history, i.e. N-G fermion superons are created
(i.e. pre-big bang is ignited) by the spontaneous breakdown of the global supertranslation
of the (super)spacetime at(above) the Planck energy and simultaneously SO(10) SP invaiant superon composite
${\it massless}$ eigenstates are spanned (gravitationally) which leads to the big bang of the universe.     

Now we summerize the results as follows. We have presented an attempt to solve the problems:
Where is SUSY realized ?  Where has N-G fermion degree of freedom gone ?
The beautiful, although qualitative, complimentality
between the gauge unified models(SM and GUTs) and SGM, i.e. the former is
strengthened or revived by taking account of the topology of the latter
superon diagram, while drawing the superon diagram of the latter is guided
by the Feynman diagram  of the former, may be an indication of SO(10) SP structure of spacetime and matter behind
the gauge models, which leads to a speculation, SGM for spacetime and matter.
Among the survived (low energy) new eigenstates of SGM, colorless eigenstates such as  
a new spin ${3 \over 2}$ lepton-type doublet
$( \nu_{\Gamma}, \Gamma^{-} )$ with the mass of the electroweak scale$( \leq Tev)$, 
a new gauge boson S, an electroweak-singlet scalar boson $(Q_{a}Q_{a}^{*}Q_{m}Q_{m}^{*})$ and 
doubly charged heavy leptons[4] $E^{--}$$({\varepsilon}_{abc}Q_{a}Q_{b}Q_{c}{\varepsilon}_{mn}Q_{m}^{*}Q_{n}^{*})$ 
are the predictions which can be tested in the near future. 
S gauge boson mass seems much larger
than the Z boson mass from the present experimental data of ${\tau}$ decay.
The clear signals of $( \nu_{\Gamma}, \Gamma^{-} )$ may be
${{e}^{+} + {e}^{-} \longrightarrow {\em l}^{+} +{\em l}^{-} +
{\em very}\ {\em large}\ {\em missing}\ {\em {P}_{T}}{\em {(E)}}}$.
The evidence of S boson will become clear in the (hadronic) decay of B meson.                    \\  
Despite these potential phenomenology, however as discussed above, the derivation of 
the low energy effective theory from SGM action (11) is a challenge, which is no more no-go. 
The classical exact solutions of vacuum EGRT can be now interpreted formally as those of ${w_{a}}^{\mu}(x)$ 
and $s^{\mu \nu}(x)$ of (11), which may indicate the existence of certain localized configurations of (composites of) 
$\psi(x)$ in (11) and favour  SGM scenario.

\vskip 30mm

The author would like to express gratitude to J. Wess
for his encouragement, enlightening discussions and interest in this work and for
the hospitality at University of M\"unchen and Max-Planck-Institut f\"ur Physik
(Werner Heisenberg Institut) and to M. Tsuda for usefull discussions at SIT.
Also he would like to thank W. Lang for useful discussions and the hospitality
at Karlsruhe University, T. Shirafuji and  Y. Tanii for useful discussions and
T. Saso for useful correspondences
and for their hospitality at Physics Department of Saitama University.
And he is also grateful to M. Gaul for help in preparing  the final manuscript.
This work is carried out under the generous auspice of  Alexander
von Humboldt Foundation, D.F.G. and in part High-Tech research program of
Saitama Institute of Technology. The author would like to thank the second referee 
for the very useful comments.

\newpage

%
\newcommand{\NP}[1]{{\it Nucl.\ Phys.\ }{\bf #1}}
\newcommand{\PL}[1]{{\it Phys.\ Lett.\ }{\bf #1}}
\newcommand{\CMP}[1]{{\it Commun.\ Math.\ Phys.\ }{\bf #1}}
\newcommand{\MPL}[1]{{\it Mod.\ Phys.\ Lett.\ }{\bf #1}}
\newcommand{\IJMP}[1]{{\it Int.\ J. Mod.\ Phys.\ }{\bf #1}}
\newcommand{\PR}[1]{{\it Phys.\ Rev.\ }{\bf #1}}
\newcommand{\PRL}[1]{{\it Phys.\ Rev.\ Lett.\ }{\bf #1}}
\newcommand{\PTP}[1]{{\it Prog.\ Theor.\ Phys.\ }{\bf #1}}
\newcommand{\PTPS}[1]{{\it Prog.\ Theor.\ Phys.\ Suppl.\ }{\bf #1}}
\newcommand{\AP}[1]{{\it Ann.\ Phys.\ }{\bf #1}}


\begin{thebibliography}{100}

\bibitem{wz}  J. Wess and B. Zumino, {\it Phys. Lett.} {\bf B49}, 52(1974).
\bibitem{GELL} M. Gell-Mann, P. Ramond and R. Slansky, Proceeding of supergravity
           workshop at Stony Brook, eds. P. van Nieuwenhuisen and
           D. Z. Freedman(North Holland, Amsterdam,1977).
\bibitem{CMHLS} S. Coleman and J. Mandula, {\it Phys. Rev.} {\bf159},1251(1967). \\
                R. Haag, J. Lopuszanski and M. Sohnius, {\it Nucl. Phys.} {\bf B88},
                257(1975).\\
                K. Shima, {\it Phys. Lett.} {\bf B312},121(1993).
\bibitem{ks1} K. Shima, {\it Z. Phys. } {\bf C18}, 25 (1983).
\bibitem{ks2} K. Shima, {\it European. Phys. J.} {\bf C7}, 341(1999). \\
              K. Shima, Proceeding of the XXth ICGTMP98 at Hobart, Australia, eds. C. Delburgo et al.
              (World Scientific, Singapore,1998).
\bibitem{gg}  H. Georgi and S. L. Glashow, {\it Phys. Rev. Lett.}, {\bf 32}, 32(1974).
\bibitem{CL} M. Artsuo et al. (CLEO Collabolation), preprint CONF98-20.
\bibitem{SK} T. Kajita(Super-Kamiokande Collabolation), \\
             Plenary talk given at the XVIIth International Conference on Neutrino
             Physics and Astrophysics, Takayama,1998.
\bibitem{va} D.V. Volkov and V.P. Akulov, {\it Phys. Lett.} {\bf B46}, 109(1973).
\bibitem{ks3} K. Shima, {\it Phys. Rev.} {\bf D{20}}, 574(1979).
              K. Shima, {\it Phys. Rev.} {\bf D{15}}, 2165(1977).
\bibitem{bv} W. Bardeen and V. Visnjic,{\it Nucl. Phys.} {\bf B194}, {422}(1982).
\bibitem{st} K. Shima and M. Tsuda, in preparation.
\bibitem{rikuzw}  M. Ro\v{c}ek, {\it Phys. Rev. Lett.} {\bf 41}, 451(1978).\\
                 E. A. Ivanov and A.A. Kapustnikov, {\it J. Phys.} 2375(1978).\\
              T. Uematsu and C.K. Zachos, {\it Nucl. Phys.} {\bf B201}, {254}(1982).  \\
              J. Wess, Karlsruhe University preprint(Festschrift for  J. Lopszanski, \\
              December,1982).
\bibitem{ws} U. Lindstr\"om and M. M. Ro\v{c}ek, \PR{D19}, 2300(1979).    \\
             S. Samuel and J. Wess,{\it Nucl. Phys.} {\bf B221}, {153}(1983).
\bibitem{ks5}  K. Shima, Plenary talk at the 3rd International Conference on Symmetry in
               Nonlinear Mathematical Physics, Kiev, Ukraine(1999). To appear in the proceeding.
\bibitem{isswz}C.J. Isham, A. Salam and J. Strathdee, {\it Phys. Rev.}, {\bf D3},
               867(1971).                                             \\
               J. Wess and B. Zumino, Brandeis Lecture, 1971.                                                              \bibitem{gn} M. Gell-Mann and Y. Ne'eMann in {\em Eightfold Way}(W.A.Benjamin,
                 N.Y.,1964).

%
\end{thebibliography}
\end{document}